\documentclass[11pt]{article}
\usepackage{amsmath}
\usepackage{amssymb}


\begin{document}

\title{A description of $1/4$ BPS configurations in minimal type IIB SUGRA}

\author{Aristomenis Donos\\
Physics Department\\
Brown University\\
Providence, Rhode Island 02912, USA}

\maketitle
\begin{abstract}
In this paper we present an effort to extend the LLM construction
of $1/2$ BPS states in minimal IIB supergravity to configurations
that preserve $1/4$ of the total number of supersymmetries. Following
the same techniques we reduce the problem to that of a single scalar
which satisfies a  non-linear equation. In particular, the scalar
is identified to be the Kahler potential with which a four dimensional
base space is equipped.\newpage{}
\end{abstract}

\section{Introduction}

In supergravity, classical solutions that break some degree of supersymmetry
have been studied extensively \cite{Lunin:2002iz,Gauntlett:2002nw,Gutowski:2003rg,Gauntlett:2004zh,Lin:2004nb}.
A particularly elegant study was given for type IIB $1/2$ BPS in
\cite{Lin:2004nb}, where the phase space of non-relativistic fermions
was shown to emerge. Not only the conserved charges, angular momentum
and five form flux were shown to agree with energy and particle number
but also the equivalence of the full dynamics was demonstrated in
\cite{Grant:2005qc,Maoz:2005nk}. The fermionic description of $1/2$
BPS states in $\mathcal{N}=4$ SYM was obtained from matrix model
reduction in \cite{Corley:2001zk,Berenstein:2004kk}. The reduced
matrix model and free fermion picture gave a successful description
of giant graviton configurations \cite{McGreevy:2000cw,Grisaru:2000zn,Hashimoto:2000zp,Bena:2004qv,Balasubramanian:2001nh,Corley:2001zk,Das:2000st,Das:2000fu,deMelloKoch:2004ws}.

The next step in constructing the full map between IIB string theory
with $AdS_{5}\times S^{5}$ asymptotics and $\mathcal{N}=4$ SYM theory
is the consideration of the map between less supersymmetric states
on both sides. On the field theory $1/4$ BPS states have been constructed
in \cite{Ryzhov:2001bp,Hoker:2001bq,Hoker:2003vf}, while even less
supersymmetric brane configurations were considered in \cite{Mikhailov:2000ya}
and more recently in \cite{Biswas:2006tj,Mandal:2006tk}. By reducing
the number of preserved of supersymmetries by a factor of 2 one is
able to double the dimensionality of the phase space in which the
states under consideration live.

Guided by the analysis of \cite{Lin:2004nb}, where a two dimensional
phase space was explicitly shown, we present an effort to enlarge
the number of deegres of freedom by reducing the number of preserved
supersymmetries to 8. Unfortunately the equation that we are called
to solve in the end of our analysis is highly non-trivial and this
is where we are forced to stop.

In the begining of the section 2 we present our ansatz for the metric
and the self dual five-form which has an$SO\left(4\right)\times SO\left(2\right)$
symmetry. We then proceed to the dimensional reduction of the Killing
spinor equation which leads to a differential equation and two projection
equations for the Killing spinor. In the following subsection we study
the constraints that the background geometry has to obey in order
for the Killing spinor to exist. In the last section we argue that
Einstein's equations are guaranteed to be satisfied, after having
checked the Bianchi identities for the five-form, as a consequence
of the integrability of the Killing spinor equation. We have also
included two appendices that the reader might find helpful where,
among other things, we list the Fierz identities that we used in the
main text.\newpage{}

\section{Geometries preserving $1/4$ of the supersymmetries}

The method that we will follow for studying the constraints imposed
on the geometry by supersymmetry was originally developed in \cite{Gauntlett:2002sc,Gauntlett:2002nw,Gutowski:2003rg,Gauntlett:2004zh,Gauntlett:2004yd}.
Demanding the existence of a Killing spinor and constructing bilinear
tensors from that spinor one can find first order equations that relate
the fluxes to the metric and also impose constraints on the metric.
Studying then the integrability conditions for the Killing spinor
equation one can derive a set of field equations that the fluxes and
the background geometry obey \cite{Gran:2005ct}.

\subsection{Reduction of the Killing spinor equation}

Having in mind the extension of the LLM geometries \cite{Lin:2004nb}
to include one more angular quantum number we make the following $SO(4)\times SO(2)$
symmetric ansatz for the metric and self-dual five-form field of minimal
IIB SUGRA in ten dimensions\begin{align}
ds^{2} & =g_{\mu\nu}dx^{\mu}dx^{\nu}+e^{H+G}d\hat{\Omega}_{3}^{2}+e^{H-G}d\psi^{2}\nonumber \\
F & =\hat{F}_{\mu_{1}\mu_{2}}dx^{\mu_{1}}\wedge dx^{\mu_{2}}\wedge d\hat{\Omega}_{3}+\tilde{F}_{\mu_{1}\cdots\mu_{4}}dx^{\mu_{1}}\wedge\cdots\wedge dx^{\mu_{4}}\wedge d\psi.\label{anz}\end{align}
From the self-duality of the five form $F=\star_{10}F$ and the Bianchi
identity we obtain the relations\begin{align}
\hat{F} & =\frac{4!}{2!}e^{2G+H}\star_{6}\tilde{F}\label{Fduality}\\
d\tilde{F} & =0\nonumber \\
d\hat{F} & =0\nonumber \end{align}
where the Hodge duality in \eqref{Fduality} is meant to be taken
with respect to the six dimensional metric $g_{\mu\nu}$ that appears
in \eqref{anz}.

Following \cite{Lin:2004nb}, the first step in the procedure is to
dimensionally reduce the ten dimensional Killing spinor equation \begin{equation}
\mathcal{D_{M}}\eta=\nabla_{M}\eta+\frac{\imath}{480}\Gamma^{M_{1}\ldots M_{5}}F_{M_{1}\ldots M_{5}}\Gamma_{M}\eta=0.\label{killingequation}\end{equation}
The product of this procedure will be a differential equation for
the Killing spinor in six dimensions and two algebraic equations that
come from the reductions on $S^{3}$ and $S^{1}$ respectively. We
decompose the ten dimensional Dirac matrices with Lorentz indices
as\begin{align*}
\Gamma_{\mu} & =\gamma_{\mu}\otimes\hat{\sigma}_{1}\otimes\mathbb{I}_{2},\quad\mu=1\ldots6\\
\Gamma_{\hat{\mu}} & =\mathbb{I}_{8}\otimes\hat{\sigma}_{2}\otimes\sigma_{\hat{\mu}},\quad\hat{\mu}=7,8,9\\
\Gamma_{10} & =\gamma_{7}\otimes\hat{\sigma}_{1}\otimes\mathbb{I}_{2},\,\gamma_{7}=\gamma_{1}\ldots\gamma_{6}\end{align*}
and the Weyl Killing spinor as\[
\eta=\epsilon\otimes\left[\begin{array}{c}
1\\
0\end{array}\right]\otimes\chi_{\alpha}=\varepsilon\otimes\chi_{\alpha}\]
where $\epsilon$ is an 8 component spinor on which the six dimensional
gamma matrices$\gamma_{\mu}$ act and $\chi_{\alpha}$ is a Killing
spinor on a sphere of unit radius and it satisfies \[
\hat{\nabla}_{\hat{a}}\chi_{\alpha}=\frac{i\alpha}{2}\sigma_{\hat{a}}\chi_{\alpha},\qquad a=\pm1.\]
Under the above decomposition the ten dimensional Weyl condition reads\[
\Gamma_{11}\eta=\Gamma_{1}\ldots\Gamma_{10}\eta=\mathbb{I}_{8}\otimes\hat{\sigma}_{3}\otimes\mathbb{I}_{2}\eta=\eta\Rightarrow\hat{\sigma}_{3}\eta=\eta.\]
For the dependence on the coordinates we have\[
\varepsilon\left(x^{\mu},\hat{\Omega}_{3},\psi\right)=e^{\frac{\imath}{2}n\psi}\varepsilon\left(x^{\mu},\hat{\Omega}_{3}\right)\]
which gives\[
\imath\partial_{\psi}\varepsilon=-\frac{n}{2}\varepsilon.\]
It is useful to rewrite the second term in \eqref{killingequation}
as\begin{align*}
M & =\frac{\imath}{480}\Gamma^{M_{1}\ldots M_{5}}F_{M_{1}\ldots M_{5}}\\
 & =\frac{\imath}{480}\left[10\Gamma^{M_{1}M_{2}}\hat{F}_{M_{1}M_{2}}e^{-\frac{3}{2}\left(G+H\right)}\Gamma^{\hat{a}\hat{b}\hat{c}}\varepsilon_{\hat{a}\hat{b}\hat{c}}-5\Gamma^{M_{1}M_{2}M_{3}M_{4}}\Gamma^{\psi}e^{-\frac{1}{2}\left(H-G\right)}\tilde{F}_{M_{1}M_{2}M_{3}M_{4}}\right]\\
 & =-\frac{1}{8}\not\hat{F}\hat{\sigma}_{2}\left(1-\hat{\sigma}_{3}\right)\end{align*}
where we used the duality \eqref{Fduality} and the duality \eqref{gammaduality}
for the six dimensional gamma matrices. We note that the spin connection
components that will be used in the reduction are given by\begin{align*}
\omega_{\hat{\mu}m\hat{\nu}} & =-\frac{1}{2}e^{\frac{1}{2}\left(H+G\right)}\hat{e}_{\hat{\mu}\hat{n}}e_{m}^{\gamma}\partial_{\gamma}\left(H+G\right)\\
\omega_{\psi m\psi} & =-\frac{1}{2}e^{\frac{1}{2}\left(H-G\right)}e_{m}^{\gamma}\partial_{\gamma}\left(H-G\right).\end{align*}
The covariant derivatives on $S^{3}$ and $S^{1}$ are then given
by the expressions \begin{align*}
\nabla_{\hat{\mu}} & =\hat{\nabla}_{\hat{\mu}}+\frac{1}{4}\Gamma_{\hat{\mu}}\Gamma^{\lambda}\partial_{\lambda}\left(G+H\right)\\
\nabla_{\psi} & =\partial_{\psi}+\frac{1}{4}\Gamma_{\psi}\Gamma^{\lambda}\partial_{\lambda}\left(H-G\right).\end{align*}
where $\hat{\nabla}$ denotes the covariant derivative on the three
sphere of unitary radius. Having collected all the necessary ingreedients
for the reduction we obtain a differential equation and two projection
conditions for the six dimensional spinor $\varepsilon$\begin{align}
\mathcal{D}_{\mu}\epsilon & =\nabla_{\mu}\varepsilon-\imath N\gamma_{\mu}\varepsilon=0\label{kileq}\\
\mathcal{D}_{S^{3}}\epsilon & =\left[\frac{\imath\alpha}{2}e^{-\frac{1}{2}\left(H+G\right)}-\frac{i}{4}\gamma^{\lambda}\partial_{\lambda}\left(H+G\right)+N\right]\varepsilon=0\label{kilS3}\\
\mathcal{D}_{\psi}\epsilon & =\left[\frac{\imath n}{2}e^{-\frac{1}{2}\left(H-G\right)}+\frac{1}{4}\gamma_{7}\gamma^{\lambda}\partial_{\lambda}\left(H-G\right)-\imath\gamma_{7}N\right]\varepsilon=0\label{kilS1}\end{align}
where\[
N=-\frac{1}{4}\not\hat{F}e^{-\frac{3}{2}\left(G+H\right)}.\]
We observe that the form of our equations looks very similar to the
ones that appear in \cite{Liu:2004ru} in the same context of reduction.
It is convenient to linearly combine the two projectors \eqref{kilS3}
and \eqref{kilS1} to obtain the equivalent ones\begin{align}
\mathcal{D}_{H}\varepsilon & =\left(\mathcal{D}_{S^{3}}-\imath\gamma_{7}\mathcal{D}_{\psi}\right)\varepsilon=\left[\frac{\imath\alpha}{2}e^{-\frac{1}{2}\left(H+G\right)}+\frac{n}{2}e^{-\frac{1}{2}\left(H-G\right)}\gamma_{7}-\frac{i}{2}\gamma^{\lambda}\partial_{\lambda}H\right]\varepsilon\label{kil1}\\
\mathcal{D}_{G}\varepsilon & =\left(\mathcal{D}_{S^{3}}+\imath\gamma_{7}\mathcal{D}_{\psi}\right)\varepsilon=\left[\frac{\imath\alpha}{2}e^{-\frac{1}{2}\left(H+G\right)}-\frac{n}{2}e^{-\frac{1}{2}\left(H-G\right)}\gamma_{7}-\frac{i}{2}\gamma^{\lambda}\partial_{\lambda}G+2N\right]\varepsilon.\label{kil2}\end{align}
As one can observe from \eqref{kileq}, when computing the covariant
derivative of any bilinear constructed from $\varepsilon$ the result
will contain the flux. We have, in many cases, found it usefull to
use \eqref{kil2} in order to eliminate the appearance of the flux
from the computation of exterior derivatives of the bilinears. Equation
\eqref{kil1} gives us linear relations between bilinears of different
ranks.

\subsection{Geometry constraints implied by supersymmetry}

As pointed out earlier the study of equations \eqref{kileq}-\eqref{kilS1}
is made simpler and more transparent with the introduction of bilinear
forms which we list below\begin{align}
f_{1} & =\bar{\varepsilon}\gamma_{7}\varepsilon\label{f1}\\
f_{2} & =\imath\bar{\varepsilon}\varepsilon\label{f2}\\
K_{\mu} & =\bar{\varepsilon}\gamma_{\mu}\varepsilon\label{km}\\
L_{\mu} & =\bar{\varepsilon}\gamma_{\mu}\gamma_{7}\varepsilon\label{lm}\\
Y_{\mu\lambda} & =\imath\bar{\varepsilon}\gamma_{\mu\nu}\gamma_{7}\varepsilon\label{yml}\\
V_{\mu\nu} & =\bar{\varepsilon}\gamma_{\mu\nu}\varepsilon\label{vmn}\\
\Omega_{\mu\nu\lambda} & =\imath\bar{\varepsilon}\gamma_{\mu\nu\lambda}\varepsilon.\label{omegamnl}\end{align}
and as we expect, since we are dealing with a spinor in six dimensions
we have the appearance of two 2-forms and a 3-form. An equivalent
set of (complex) bilinears, and more convenient when considering Fierz
identities, is given by\begin{align}
Z^{+} & =\frac{1}{2}\left(-f_{1}-\imath f_{2}\right)=\bar{\varepsilon_{+}}\varepsilon_{-}\label{Z+}\\
Z^{-} & =\frac{1}{2}\left(f_{1}-\imath f_{2}\right)=\bar{\varepsilon_{-}}\varepsilon_{+}\label{Z-}\\
l_{\mu}^{+} & =\frac{1}{2}\left(L_{\mu}+K_{\mu}\right)=\bar{\varepsilon_{+}}\gamma_{\mu}\varepsilon_{+}\label{l+}\\
l_{\mu}^{-} & =\frac{1}{2}\left(-L_{\mu}+K_{\mu}\right)=\bar{\varepsilon_{-}}\gamma_{\mu}\varepsilon_{-}\label{l-}\\
U_{\mu\nu}^{+} & =\frac{1}{2}\left(V_{\mu\nu}-\imath Y_{\mu\nu}\right)=\bar{\varepsilon_{+}}\gamma_{\mu\nu}\varepsilon_{-}\label{Up}\\
U_{\mu\nu}^{-} & =\frac{1}{2}\left(V_{\mu\nu}+\imath Y_{\mu\nu}\right)=\bar{\varepsilon_{-}}\gamma_{\mu\nu}\varepsilon_{+}\label{Um}\\
q_{\mu\nu\lambda}^{\pm} & =\imath\bar{\varepsilon_{\pm}}\gamma_{\mu\nu\lambda}\varepsilon_{\pm}\label{q+-}\end{align}
where as usually\begin{align*}
\varepsilon_{\pm} & =\frac{1}{2}\left(\mathbb{I}_{8}\pm\gamma_{7}\right)\varepsilon\\
\gamma_{7}\varepsilon_{\pm} & =\pm\varepsilon_{\pm}.\end{align*}
Because of the duality relation \eqref{gammaduality} that the Dirac
matrices satisfy, the three form $q^{+}$ is self-dual while the three
form $q^{-}$ is antiself-dual.

Using the differential equation \eqref{kileq} one can obtain the
differential identities that govern the forms listed in \eqref{f1}-\eqref{omegamnl}\begin{align}
\nabla_{\mu}f_{1} & =\frac{\imath}{4}\bar{\varepsilon}\left(\gamma_{\mu}\gamma_{\kappa\lambda}\gamma_{7}+\gamma_{\kappa\lambda}\gamma_{\mu}\gamma_{7}\right)\varepsilon\, F^{\kappa\lambda}e^{-\frac{3}{2}\left(G+H\right)}\nonumber \\
 & =\frac{\imath}{2\cdot3!}\varepsilon_{\mu\kappa\lambda\rho\sigma\tau}\bar{\varepsilon}\left(\gamma^{\rho\sigma\tau}\right)\varepsilon\, F^{\kappa\lambda}e^{-\frac{3}{2}\left(G+H\right)}\nonumber \\
 & =\frac{1}{2\cdot3!}\varepsilon_{\mu\kappa\lambda\rho\sigma\tau}\Omega^{\rho\sigma\tau}F^{\kappa\lambda}e^{-\frac{3}{2}\left(G+H\right)}\nonumber \\
 & =\frac{1}{2}\star\left(F\wedge\Omega\right)_{\mu}e^{-\frac{3}{2}\left(G+H\right)}.\label{df1}\\
 & =-\frac{1}{3!}F_{\mu\rho\sigma\tau}\Omega^{\rho\sigma\tau}e^{\frac{1}{2}\left(G-H\right)}\end{align}
\begin{align}
\nabla_{\mu}f_{2} & =\frac{1}{2}\bar{\varepsilon}\left(\gamma_{\lambda}g_{\mu\kappa}-g_{\mu\lambda}\gamma_{\kappa}\right)\varepsilon\, F^{\kappa\lambda}e^{-\frac{3}{2}\left(G+H\right)}\nonumber \\
 & =e^{-\frac{3}{2}\left(G+H\right)}F_{\mu\lambda}K^{\lambda}.\label{df2}\end{align}
\begin{align}
\nabla_{\mu}K_{\rho} & =-\frac{\imath}{4}F^{\kappa\lambda}e^{-\frac{3}{2}\left(G+H\right)}\bar{\varepsilon}\left(\gamma_{\rho}\gamma_{\kappa\lambda}\gamma_{\mu}-\gamma_{\mu}\gamma_{\kappa\lambda}\gamma_{\rho}\right)\varepsilon\nonumber \\
 & =e^{-\frac{3}{2}\left(G+H\right)}f_{2}F_{\mu\rho}-e^{-\frac{3}{2}\left(G+H\right)}\frac{1}{4}F^{\kappa\lambda}\varepsilon_{\mu\rho\pi\tau\kappa\lambda}Y^{\pi\tau}\label{dkm}\\
 & =e^{-\frac{3}{2}\left(G+H\right)}f_{2}F_{\mu\rho}+\frac{1}{2}e^{\frac{1}{2}\left(G-H\right)}F_{\mu\rho\pi\tau}Y^{\pi\tau}\end{align}
\begin{align}
\nabla_{\mu}L_{\rho} & =\frac{\imath}{4}F^{\kappa\lambda}e^{-\frac{3}{2}\left(G+H\right)}\bar{\varepsilon}\left(\gamma_{\rho}\gamma_{\kappa\lambda}\gamma_{\mu}+\gamma_{\mu}\gamma_{\kappa\lambda}\gamma_{\rho}\right)\gamma_{7}\varepsilon\nonumber \\
 & =e^{-\frac{3}{2}\left(G+H\right)}\left[F_{\mu}^{\:\lambda}Y_{\lambda\rho}+F_{\rho}^{\:\lambda}Y_{\lambda\mu}+\frac{1}{2}g_{\mu\rho}F^{\kappa\lambda}Y_{\kappa\lambda}\right]\label{dlm}\end{align}
\begin{equation}
\nabla_{\gamma}V_{\delta\epsilon}=-e^{-\frac{3}{2}\left(G+H\right)}\left[g_{\gamma\left[\epsilon\right.}\Omega_{\left.\delta\right]\alpha\beta}F^{\alpha\beta}-F_{\gamma}^{\:\beta}\Omega_{\beta\delta\epsilon}+2\Omega_{\alpha\gamma\left[\delta\right.}F_{\:\left.\epsilon\right]}^{\alpha}\right].\label{dvmn}\end{equation}
As one expects we have the appearance of higher rank tensors on the
right hand sides of the differential equations for the bilinears when
comparing to the analysis of \cite{Lin:2004nb}. We now take a derivative
of \eqref{omegamnl} giving us\begin{align*}
\nabla_{\kappa}\Omega_{\mu\nu\lambda} & =-\bar{\varepsilon}\left(\gamma_{\mu\nu\lambda}N\gamma_{\kappa}-\gamma_{\kappa}N\gamma_{\mu\nu\lambda}\right)\varepsilon\\
 & =\frac{1}{4}e^{-\frac{3}{2}\left(G+H\right)}F^{\pi\rho}\bar{\varepsilon}\left(\gamma_{\mu\nu\lambda}\gamma_{\pi\rho}\gamma_{\kappa}-\gamma_{\kappa}\gamma_{\pi\rho}\gamma_{\mu\nu\lambda}\right)\varepsilon.\end{align*}
After antisymmetrization we have that \begin{equation}
d\Omega_{\kappa\lambda\mu\nu}=4f_{1}e^{-\frac{1}{2}\left(H-G\right)}\tilde{F}_{\kappa\lambda\mu\nu}.\label{dOmegaF}\end{equation}
The above equation, as we will see later after fixing the form of
$f_{1}$, fixes the 4-form and it gives us the Bianchi equation for
it. One can then use the duality \eqref{Fduality} to determine the
2-form $F_{\mu\nu}$, which at this point is not obvious why it will
satisfy the Bianchi identity. For the dual 3-form we have the equation\begin{align*}
\nabla_{\kappa}\left(\star\Omega\right)_{\mu\nu\lambda} & =-\bar{\varepsilon}\left(\gamma_{\mu\nu\lambda}\gamma_{7}N\gamma_{\kappa}-\gamma_{\kappa}N\gamma_{\mu\nu\lambda}\gamma_{7}\right)\varepsilon\\
 & =-\frac{1}{4}e^{-\frac{3}{2}\left(G+H\right)}F^{\pi\rho}\bar{\varepsilon}\left(\gamma_{\mu\nu\lambda}\gamma_{\pi\rho}\gamma_{\kappa}+\gamma_{\kappa}\gamma_{\pi\rho}\gamma_{\mu\nu\lambda}\right)\gamma_{7}\varepsilon.\end{align*}
Antisymmetrizing the last equation in $\kappa,\mu,\nu,\lambda$ we
obtain\begin{equation}
\left(d\star\Omega\right)_{\kappa\mu\nu\lambda}=2\bar{\varepsilon}\left(\gamma_{\mu\nu\lambda\kappa}N+N\gamma_{\kappa\mu\nu\lambda}\right)\gamma_{7}\varepsilon\label{d*omega}\end{equation}
which as we see later constrains a 4-dimensional submanifold to a
Kahler manifold.

The immidiate consequences of the vector bilinears that we formed
is the proof of the existence of a Killing vector for the six dimensional
metric in \eqref{anz} and a closed form. The above can be seen by
considering the symmetric part of \eqref{dkm} which gives\[
\nabla_{\left(\mu\right.}K_{\left.\nu\right)}=0\]
and the antisymmetric part of \eqref{dlm} which leads to\[
\nabla_{\left[\mu\right.}L_{\left.\nu\right]}=0\Rightarrow dL=0.\]
As we show in the Appendix the vectors $l^{\pm}$ are null and as
a consequence we have that \begin{align}
L^{2} & =-K^{2}=2\, l^{+}\cdot l^{-}\label{L2K2}\\
K\cdot L & =0.\label{KL}\end{align}
We now follow an argument presented in \cite{Gutowski:2003rg} applied
to our chiral spinors. As we prove in the Appendix using Fierz identities
we have the following relations\begin{align*}
i_{l^{+}}q^{+} & =0\\
i_{l^{-}}q^{-} & =0\end{align*}
and the dualities of $q^{\pm}$ imply that\[
l^{\pm}\wedge\, q^{\pm}=0.\]
After the above observations we conclude that in a coordinate system
where the metric takes the form\begin{equation}
ds^{2}=e^{+}e^{-}+\delta_{ab}e^{a}e^{b},\qquad a,b=1\ldots4\label{metric}\end{equation}
where\begin{align*}
l^{+} & =e^{+}\\
l^{-} & =e^{-}\end{align*}
the two three forms can be written as\begin{align}
q^{+} & =l^{+}\wedge\, I\label{Idef}\\
q^{-} & =l^{-}\wedge\, J\label{Jdef}\end{align}
where the two forms\begin{align}
I & =\frac{1}{2}I_{ab}e^{a}\wedge\, e^{b}\label{Idef2}\\
J & =\frac{1}{2}J_{ab}e^{a}\wedge\, e^{b}.\label{Jdef2}\end{align}
are anti-selfdual with respect to the metric\begin{equation}
ds^{2}=\delta_{ab}e^{a}e^{b},\qquad a,b=1\ldots4\label{4metric}\end{equation}
with orientation defined by $\varepsilon^{+-abcd}=\varepsilon^{abcd}$.
From equations \eqref{IIpr}-\eqref{id3} one can also prove that
\begin{align}
I_{\: b}^{a}I_{\: c}^{b} & =-\delta_{c}^{a}\label{II}\\
J_{\: b}^{a}J_{\: c}^{b} & =-\delta_{c}^{a}.\nonumber \end{align}
These two equation imply that that the 2-forms $I$ and $J$ consist
complex structures for the metric \eqref{4metric} which as we will
later prove are in fact equal rendering the four dimensial manifold
with metric $\delta_{ab}e^{a}e^{b}$ pre-Kahler. From the above considerations
we see that \begin{align}
q^{+\pi\rho\sigma}q_{\pi\rho\sigma}^{-} & =3\left[\left(l^{+}\cdot l^{-}\right)\left(I^{ab}J_{ab}\right)+2l^{+m}l^{-n}I_{nk}J_{\; m}^{k}\right]\nonumber \\
 & =3\left(l^{+}\cdot l^{-}\right)\left(I^{ab}J_{ab}\right)\label{q+q-}\end{align}
which in combination with \eqref{ZZ} can give us the product $l^{+}\cdot l^{-}$in
terms of the functions $Z^{+}$ and $Z^{-}$, we will come back to
this later.

We can combine equation \eqref{df1} with \eqref{v3} and \eqref{df2}
with \eqref{v4} to relate $f_{1}$ and $f_{2}$ to $G$ and $H$
as\begin{align}
\partial_{\mu}f_{1} & =\frac{1}{2}f_{1}\partial_{\mu}\left(H-G\right)\Rightarrow\label{df12}\\
f_{1} & =\lambda e^{\frac{1}{2}\left(H-G\right)}\nonumber \end{align}
and\begin{align}
\partial_{\mu}f_{2} & =\frac{1}{2}f_{2}\partial_{\mu}\left(H+G\right)\Rightarrow\label{df22}\\
f_{2} & =\kappa e^{\frac{1}{2}\left(H+G\right)}.\nonumber \end{align}
Combining \eqref{df1} with \eqref{v1} and \eqref{df2} with \eqref{v2}
we obtain the relation\begin{equation}
\partial_{\mu}e^{H}=\frac{n}{\kappa}L_{\mu}\label{H}\end{equation}
and the constrain\begin{equation}
\frac{\alpha}{\lambda}=-\frac{n}{\kappa}.\end{equation}
Adding equations \eqref{omega1} and \eqref{omega2} we obtain\[
\Omega_{\mu\nu\lambda}\partial^{\lambda}H=0\]
which in combination with \eqref{H} yields\begin{align}
i_{L}\Omega & =0\Rightarrow\label{LOmega=0}\\
i_{l^{+}}q^{-} & =i_{l^{-}}q^{+}\label{lqlq}\end{align}
where we used \eqref{l+q+} and \eqref{l-q-}. Equation \eqref{lqlq}
and the equation\[
i_{l^{\pm}}I=i_{l^{\pm}}J=0\]
helps us relate the two 2-forms as\begin{equation}
I=J.\label{IJ}\end{equation}
which is one of the supersymmetry requirements following from the
Killing spinor equation. After this observation we may write\begin{align}
\Omega & =K\wedge I\label{OKI}\\
\star\Omega & =L\wedge I\label{OLI}\end{align}
and from equations \eqref{II} and \eqref{q+q-} we have that\[
q^{+\pi\rho\sigma}q_{\pi\rho\sigma}^{-}=12\, l^{+}\cdot l^{-}.\]
We now see from the Fierz identity\eqref{ZZ} that we did a little
more work to recover the familiar result from \cite{Lin:2004nb} \[
L^{2}=-K^{2}=f_{1}^{2}+f_{2}^{2}.\]
We now turn our attention to the algebraic relation \eqref{Omega_wedge_L}
which we will use to express the complex structure $I$ in terms of
other bilinears. For this reason we contract equation \eqref{OLI}
with the vector $L^{\mu}$ to obtain\[
I=\frac{1}{f_{1}^{2}+f_{2}^{2}}\, i_{L}\star\Omega.\]
We may now use equations \eqref{l+U+}, \eqref{l-U-}, \eqref{l+U-}
and \eqref{l-U+} to express \[
i_{L}V=f_{1}K.\]
Contracting equation equation \eqref{Omega_wedge_L} with $L^{\mu}$
yields a relation between the complex structure and the 2-form constructed
from bilinears \[
I=-\frac{1}{k}e^{-\frac{1}{2}\left(H+G\right)}V+\frac{f_{1}e^{-\frac{1}{2}\left(G+H\right)}}{f_{1}^{2}+f_{2}^{2}}L\wedge K.\]
Having in mind the closure of the complex structure we are tempted
to consider the external derivative of the above equation which in
turn gives \begin{equation}
dI=\frac{1}{2k}e^{-\frac{1}{2}\left(G+H\right)}d\left(G+H\right)\wedge V-\frac{1}{k}e^{-\frac{1}{2}\left(G+H\right)}dV-\frac{f_{1}e^{-\frac{1}{2}\left(G+H\right)}}{f_{1}^{2}+f_{2}^{2}}L\wedge dK+L\wedge K\wedge d\left[\frac{f_{1}e^{-\frac{1}{2}\left(G+H\right)}}{f_{1}^{2}+f_{2}^{2}}\right].\label{dI1}\end{equation}
In order to evaluate $dV$ we go back to \eqref{dvmn} and we observe
that\begin{align*}
dV_{\kappa\mu\nu} & =-\frac{3\imath}{4}e^{-\frac{3}{2}\left(G+H\right)}F^{\pi\rho}\bar{\varepsilon}\left(\gamma_{\left[\mu\nu\right.}\gamma_{\pi\rho}\gamma_{\left.\kappa\right]}-\gamma_{\left[\kappa\right.}\gamma_{\pi\rho}\gamma_{\left.\mu\nu\right]}\right)\varepsilon\\
 & =-\frac{3\imath}{12}e^{-\frac{3}{2}\left(G+H\right)}F^{\pi\rho}\bar{\varepsilon}\left(\gamma_{\kappa\mu\nu}\gamma_{\pi\rho}-\gamma_{\pi\rho}\gamma_{\kappa\mu\nu}\right)\varepsilon\\
 & =\imath\bar{\varepsilon}\left(\gamma_{\kappa\mu\nu}N-N\gamma_{\kappa\mu\nu}\right)\varepsilon.\end{align*}
As we described before we will exploit that the right hand of the
previous equation take and we will use \eqref{kil2} to relate the
derivative of $V$ to fields that do not include the flux \[
dV=-\frac{1}{2}V\wedge dG+\frac{n}{2}e^{-\frac{1}{2}\left(H-G\right)}L\wedge I.\]
At this point we pause our analysis of \eqref{dI1} in order to fix
our gauge. As in \cite{Lin:2004nb} we use the closed form $L$ to
identify a coordinate which we also call $y$ with the analogous geometrical
meaning of the product of the radii of $S^{3}$ and $S^{1}$that appear
in our ansatz \eqref{anz}. We make the gauge choice\begin{align*}
e^{+} & =X\left(dt+A_{m}dx^{m}\right)+Bdy\\
e^{-} & =-X\left(dt+A_{m}dx^{m}\right)+Bdy\\
K & =-Xdt-XA\\
L & =\gamma dy\end{align*}
and from equations \eqref{L2K2} and \eqref{KL} we draw the conclusion
that\[
X=B^{-1}=h^{-2}=f_{1}^{2}+f_{2}^{2}.\]
At this point the ten dimensional metric has the form\[
ds^{2}=-\frac{1}{h^{2}}\left(dt+A\right)^{2}+h^{2}dy^{2}+e^{-G-H}h_{mn}dx^{m}dx^{n}+\frac{\gamma n}{k}ye^{G}d\hat{\Omega}_{3}^{2}+\frac{\gamma n}{k}ye^{-G}d\psi^{2},\qquad m=1,\ldots,4\]
where we used \eqref{H} to fix\[
e^{H}=\frac{\gamma n}{k}y.\]
For later convenience we have rescaled the vielbein \eqref{metric}
as\[
e_{\mu}^{a}=e^{-\frac{1}{2}G-\frac{1}{2}H}\tilde{e}_{\mu}^{a}\]
and with this choice we have defined\begin{equation}
h_{\mu\nu}=\delta_{ab}\tilde{e}_{\mu}^{a}\tilde{e}_{\nu}^{b}.\label{hmn}\end{equation}
We also rescale the complex structure accordingly and define\begin{equation}
\mathcal{J}=e^{G+H}I\label{J}\end{equation}
which now satisfies \[
\mathcal{J}_{\quad p}^{m}\mathcal{J}_{\quad n}^{p}=-\delta_{n}^{m}\]
while raising of indices in the above formula is done using \eqref{hmn}. 

We now resume the analysis of \eqref{dI1} and express the derivative
of $K$ as \[
dK=\frac{f_{2}^{2}-f_{1}^{2}}{f_{2}^{2}+f_{1}^{2}}dG\wedge K+dH\wedge K-\left(f_{2}^{2}+f_{1}^{2}\right)dA.\]
After a little algebra equation \eqref{dI1} takes the form\begin{align}
dI & =-\left(dH+dG\right)\wedge I+f_{1}e^{-\frac{1}{2}\left(G+H\right)}L\wedge dA.\label{dI}\end{align}
In terms of $\mathcal{J}$ equation \eqref{dI} reads\begin{equation}
d\mathcal{J}=f_{1}e^{\frac{1}{2}\left(G+H\right)}L\wedge dA.\label{dmathcalJ}\end{equation}
It is instructive to split the operation of external differentiation
as\[
d=\tilde{d}+d_{y}+d_{t}\]
defned as\begin{align*}
\tilde{d}f & =\partial_{x^{i}}f\wedge dx^{i},\quad i=1\dots4\\
d_{y}f & =\partial_{y}f\wedge dy\\
d_{t}f & =\partial_{t}f\wedge dt.\end{align*}
After the above spliting we observe from \eqref{dmathcalJ} that\begin{align}
\tilde{d}\mathcal{J} & =0\nonumber \\
\partial_{y}\mathcal{J} & =\frac{\gamma\lambda n}{k}y\tilde{d}A.\label{dyJ}\end{align}
From the first equation we see that the four dimensional metric equiped
with the metric \eqref{hmn} is Kahler for each $y$. We now consider
\eqref{dOmegaF} and using \[
\Omega=K\wedge I\]
we have that the four form is given by\begin{align*}
F & =\frac{e^{\frac{1}{2}\left(H-G\right)}}{4f_{1}}d\Omega\\
 & =\frac{e^{\frac{1}{2}\left(H-G\right)}}{4f_{1}}\left[dK\wedge I-K\wedge dI\right]\\
 & =\frac{e^{\frac{1}{2}\left(H-G\right)}}{4f_{1}}\left[\frac{f_{2}^{2}-f_{1}^{2}}{f_{2}^{2}+f_{1}^{2}}\, dG\wedge K\wedge I+\: dH\wedge K\wedge I-\left(f_{2}^{2}+f_{1}^{2}\right)\: dA\wedge I\right]\\
 & \quad-\frac{e^{\frac{1}{2}\left(H-G\right)}}{4f_{1}}K\wedge\left[-\left(dG+dH\right)\wedge I+f_{1}e^{-\frac{1}{2}\left(G+H\right)}\: K\wedge L\wedge dA\right]\\
 & =\frac{e^{\frac{1}{2}\left(H-G\right)}}{4f_{1}}\left[-\frac{2f_{1}^{2}}{f_{1}^{2}+f_{2}^{2}}\, dG\wedge K\wedge I+f_{1}e^{-\frac{1}{2}\left(G+H\right)}\, L\wedge K\wedge dA-\left(f_{2}^{2}+f_{1}^{2}\right)\: dA\wedge I\right].\end{align*}
We now consider \eqref{df1} and contract it with $L^{\mu}$. This
gives us\begin{align*}
L^{\mu}\partial_{\mu}f_{1}^{2} & =-\frac{1}{4}d\Omega_{\mu\rho\sigma\tau}L^{\mu}K^{\rho}I^{\sigma\tau}\\
\gamma f_{1}^{2}\partial_{y}\left(G+H\right) & =\frac{1}{4}f_{1}e^{-\frac{1}{2}\left(G+H\right)}\left(f_{1}^{2}+f_{2}^{2}\right)\, I^{ij}\tilde{d}A_{ij}.\end{align*}
The last term in the second equation can be found using \eqref{dyJ}
and noting that\[
\mathcal{J}\wedge\mathcal{J}=-2\,\sqrt{h}\, dz\wedge d\bar{z}\wedge dw\wedge d\bar{w}\]
where\[
\sqrt{h}=\left|\begin{array}{cc}
\partial_{z}\partial_{\bar{z}}K & \partial_{w}\partial_{\bar{z}}K\\
\partial_{z}\partial_{\bar{w}}K & \partial_{w}\partial_{\bar{w}}K\end{array}\right|.\]
We also need to remember that the complex structure $\mathcal{J}$
is anti-selfdual which leads us to the conclusion\begin{align*}
-4\partial_{y}\sqrt{h} & =-2\frac{\gamma\lambda n}{k}y\sqrt{h}e^{-G-H}I^{ij}\tilde{d}A_{ij}\\
 & =-8\sqrt{h}\frac{\lambda e^{-G}f_{1}\partial_{y}\left(G+H\right)}{e^{-\frac{1}{2}\left(G+H\right)}\left(f_{1}^{2}+f_{2}^{2}\right)}\end{align*}
which in the end gives us the equation that related the volume of
the four dimensional base space to the scalar $G$\begin{align}
\partial_{y}\ln\sqrt{h} & =\frac{2\lambda^{2}e^{-2G}\partial_{y}G}{k^{2}\left(1+\frac{\lambda^{2}}{k^{2}}e^{-2G}\right)}+\frac{2\lambda^{2}e^{-2G}\partial_{y}H}{k^{2}\left(1+\frac{\lambda^{2}}{k^{2}}e^{-2G}\right)}\nonumber \\
\partial_{y}\ln\sqrt{h} & =-\partial_{y}\ln\left(1+\frac{\lambda^{2}}{k^{2}}e^{-2G}\right)+\left(1-z\right)\partial_{y}H\label{dyh}\end{align}
where we have set\[
z=\frac{1-\frac{\lambda^{2}}{k^{2}}e^{-2G}}{1+\frac{\lambda^{2}}{k^{2}}e^{-2G}}.\]

At this point we would like to see what the duality \eqref{Fduality}
has to give. We evaluate the component\begin{align*}
F_{tx} & =\frac{e^{2G+H}}{4!}\epsilon_{tx}^{\quad\pi\rho\sigma\tau}F_{\pi\rho\sigma\tau}\\
 & =\frac{e^{\frac{3}{2}\left(H+G\right)}}{4!\cdot4f_{1}}\epsilon_{tx}^{\quad\pi\rho\sigma\tau}d\Omega_{\pi\rho\sigma\tau}\\
 & =\frac{1}{8}\frac{e^{\frac{3}{2}\left(G+H\right)}}{f_{1}}\left(f_{1}^{2}+f_{2}^{2}\right)^{2}\partial_{y}A_{x^{1}}I_{x^{2}x^{3}}\epsilon_{x}^{\; x^{1}x^{2}x^{3}}\\
 & =-\frac{1}{4}\frac{e^{\frac{3}{2}\left(G+H\right)}}{f_{1}}\left(f_{1}^{2}+f_{2}^{2}\right)^{2}\partial_{y}A_{x^{1}}I_{x}^{\; x^{1}}.\end{align*}
Where we have used the anti-self duality of $I$. From \eqref{f2}
we have that\[
F_{tx}=-\frac{1}{2}e^{\frac{3}{2}\left(G+H\right)}f_{2}\partial_{x}G.\]
Equating the right hand sides of the above two equations we have\begin{align*}
I_{x}^{\; x^{1}}\partial_{y}A_{x^{1}} & =\frac{2f_{1}f_{2}}{\left(f_{1}^{2}+f_{2}^{2}\right)^{2}}\partial_{x}G\\
\partial_{y}A_{x^{1}} & =-\frac{2f_{1}f_{2}}{\left(f_{1}^{2}+f_{2}^{2}\right)^{2}}I_{x}^{\; x^{1}}\partial_{x^{1}}G\end{align*}
where we used \eqref{II}. We can rewrite the above expression in
the form\begin{align*}
\partial_{y}A_{x^{1}} & =-\frac{2f_{1}f_{2}}{\left(f_{1}^{2}+f_{2}^{2}\right)^{2}}I_{x}^{\; x^{1}}\partial_{x^{1}}G\\
 & =-2k\lambda e^{-H}\frac{1}{\left(\lambda^{2}e^{-G}+k^{2}e^{G}\right)^{2}}I_{x}^{\; x^{1}}\partial_{x^{1}}G\\
 & =-\frac{e^{-H}}{2k\lambda^{3}}I_{x}^{\; x^{1}}\partial_{x^{1}}z.\end{align*}
The last equation together with \eqref{dyJ} gives us all the non-trivial
components of $dA$. Imposing the consistency condition $d^{2}A=0$,
the $x^{1}x^{2}x^{3}$ component of the equation gives us back the
closure, whlie the $y,x^{1},x^{2}$ component gives us another relation
between the Kahler potential and the scalar $G$ \begin{equation}
2k\lambda^{2}e^{H}\partial_{y}\left(e^{-H}\partial_{y}\mathcal{J}\right)+d\left(\mathcal{J}\cdot\partial z\right)=0.\label{scalarana}\end{equation}
The last equation may be viewed as the analog of the Laplace equation
that appears in the LLM construction of $1/2$ BPS states. At this
point we have reached two equations \eqref{dyh}, \eqref{scalarana}
for two scalars, the Kahler potential and $G$ that is used to parameterize
the radii of $S^{3}$ and $S^{1}$in our ansatz \eqref{anz}. From
the definition of the complex structure we come to the conclusion
that up to a harmonic function\begin{equation}
z=-2y\partial_{y}\left(\frac{1}{y}\partial_{y}K\right).\label{zK}\end{equation}
Using the above equation we can now integrate equation \eqref{dyh}
and give an equation that governs the Kahler potential%
\footnote{A version of this equation was obtained by O. Lunin\cite{Lunin:talk}%
}\begin{equation}
\left|\begin{array}{cc}
\partial_{z}\partial_{\bar{z}}K & \partial_{w}\partial_{\bar{z}}K\\
\partial_{z}\partial_{\bar{w}}K & \partial_{w}\partial_{\bar{w}}K\end{array}\right|=y\frac{e^{\frac{2}{y}\partial_{y}K}}{2}\left(-2y\partial_{y}\left(\frac{1}{y}\partial_{y}K\right)+1\right)\label{MoAm}\end{equation}
where we caution the reader that we have chosen to have a trivial
{}``initial condition'' for the differential equation \eqref{dyh}.

It is important to check that the 2-form that one finds through \eqref{Fduality}
is closed. Here we list the components of the two form that we find
by dualizing the 4-form\begin{align*}
F_{tx} & =-\frac{1}{2}e^{\frac{3}{2}\left(G+H\right)}f_{2}\partial_{x}G\\
F_{ty} & =-\frac{1}{4}\partial_{y}e^{2\left(G+H\right)}\\
F_{yx} & =\frac{1}{4}A_{x}\partial_{y}e^{2\left(G+H\right)}-\frac{1}{8}e^{G+H}\left(e^{G}+e^{-G}\right)\,\mathcal{J}_{x}^{\; x_{1}}\partial_{x^{1}}z\\
F_{x_{1}x_{2}} & =\frac{1}{2}\mathcal{J}_{x_{1}x_{2}}-\frac{1}{4}y\partial_{y}\mathcal{J}_{x_{1}x_{2}}-\frac{1}{4}ye^{2G}\partial_{y}\mathcal{J}_{x_{1}x_{2}}-\frac{1}{4}y^{2}\partial_{\left[x_{1}\right.}e^{2G}A_{\left.x_{2}\right]}.\end{align*}
Checking then whether $dF=0$ brings us in front of equations which
we have already seen in previous considerations. During this check
we don't need to use the specific forms \eqref{zK} or \eqref{MoAm}
but only the differential equations \eqref{dyh} and \eqref{scalarana}.

\section{Einstein's field equations}

In order to check Einstein's equation one considers the integrability
conditions of the Killing spinor equation. In order to do this we
could follow the analysis of \cite{Liu:2004ru} or \cite{Liu:2004hy}
and do everything after the dimensional reduction. However, we will
follow the discussion of \cite{Gran:2005ct} directly for the ten
dimensional theory. We now present the argument here for completenes.
The solutions that we have tried to describe up to now come from the
requirement of existence of a commuting Weyl spinor for which the
supercovariant derivative vanishes\[
\mathcal{D}_{M}\eta=\nabla_{M}\eta+\frac{\imath}{480}\Gamma^{M_{1}\ldots M_{5}}F_{M_{1}\ldots M_{5}}\Gamma_{M}\eta=0.\]
Since the above relation is true we conclude that the commutator of
two supercovariant derivatives should also be zero when applied on
the same Killing spinor\[
\mathcal{R}_{MN}\eta=\left[\mathcal{D}_{M},\mathcal{D}_{N}\right]\eta=0.\]
The commutator of the supercovariant derivatives can be found in \cite{Papadopoulos:2003jk}.
Contracting the commutator with the appropriate gamma matrices gives
us the relation\begin{equation}
\frac{1}{2}\Gamma_{A}^{\; MN}\left[\mathcal{D}_{M},\mathcal{D}_{N}\right]=\frac{1}{2}E_{AM}\Gamma^{M}+\frac{\imath}{3!}\Gamma^{M_{1}M_{2}M_{3}}\nabla^{B}F_{AM_{1}M_{2}M_{3}B}\label{eeom}\end{equation}
 where\[
E_{MN}=R_{MN}-\frac{1}{2}g_{MN}R-\frac{1}{6}F_{MA_{1}A_{2}A_{3}A_{4}}F_{N}^{\; A_{1}A_{2}A_{3}A_{4}}.\]
One can relate the Bianchi identity of the five form to its equation
of motion through duality. Having this in mind and the fact that the
geometries that we described demand a closed five form we can conclude
from \eqref{eeom} that they satisfy Einstein's field equations.

\section{Conclusions and summary}

In this paper we have derived the equation that governs $1/4$ BPS
states in minimal IIB supergravity. After making an $SO\left(4\right)\times SO\left(2\right)$
symmetric ansatz for our fields we used the powerful techniques that
were developed in \cite{Gauntlett:2002sc,Gauntlett:2002nw,Gutowski:2003rg,Gauntlett:2004zh,Gauntlett:2004yd}
to find the constraints imposed on the background geometry by the
existence of a Killing spinor. During this process we showed that
every field can be solved for in terms of a single scalar function
which appears in our equations as the Kahler potential of a four dimensional
base space which is Kahler.

A very relevant future topic which needs to be understood is the physical
moduli space which parameterizes regular $1/4$ BPS solutions. This
involves study of regularity conditions which will lead to further
reduction of the space. An equally important problem is the recovery
of the same moduli space from SYM theory. It is clear that it is related
to the dynamics of more than one complex matrices.

\section*{Acknowledgements}

We would like to thank O. Lunin for the useful private discussions
and his comments on this work. We would also like to thank A. Jevicki,
R. McNees and R. de Mello Koch for many enlightening and useful discussions.

\textbf{Note added:} After the appearance of this paper we became
aware of \cite{Kim:2005ez} where the author, among other properties,
has shown an interesting six dimensional Kahler structure for the
case of BPS states containing a factor of $AdS_{3}$ (or $S^{3}$)
in type IIB supergravity.\newpage{}

\appendix

\section{Fierz Identities}

In this appendix we list various identities for the six dimensional
Clifford algebra that we used in the main text of the paper. We start
by giving the duality relation for the gamma matrices which we heavily
use \begin{equation}
\gamma^{a_{1}\ldots a_{n}}=\frac{\left(-1\right)^{\left[\frac{n}{2}\right]+1}}{\left(6-n\right)!}\varepsilon^{a_{1}\ldots a_{n}b_{b+1}\ldots b_{6-n}}\gamma_{b_{1}\ldots b_{6-n}}\gamma_{7}.\label{gammaduality}\end{equation}
The Fierz identities that we will use among the bilinear forms can
be derived by the basic formula\begin{align}
\bar{\psi_{1}}\psi_{2}\bar{\psi}_{3}\psi_{4} & =\frac{1}{8}\left[\bar{\psi_{1}}\psi_{4}\bar{\psi}_{3}\psi_{2}+\bar{\psi_{1}}\gamma_{7}\psi_{4}\bar{\psi}_{3}\gamma_{7}\psi_{2}-\frac{1}{2}\bar{\psi_{1}}\gamma_{\mu\nu}\psi_{4}\bar{\psi}_{3}\gamma^{\mu\nu}\psi_{2}-\frac{1}{2}\bar{\psi_{1}}\gamma_{\mu}\gamma_{7}\psi_{4}\bar{\psi}_{3}\gamma^{\mu\nu}\gamma_{7}\psi_{2}\right]\nonumber \\
 & +\frac{1}{8}\left[\bar{\psi_{1}}\gamma_{\mu}\psi_{4}\bar{\psi}_{3}\gamma^{\mu}\psi_{2}-\bar{\psi_{1}}\gamma_{\mu}\gamma_{7}\psi_{4}\bar{\psi}_{3}\gamma^{\mu}\gamma_{7}\psi_{2}\right]\nonumber \\
 & -\frac{1}{96}\left[\bar{\psi_{1}}\gamma_{\mu\nu\lambda}\psi_{4}\bar{\psi}_{3}\gamma^{\mu\nu\lambda}\psi_{2}-\bar{\psi_{1}}\gamma_{\mu\nu\lambda}\gamma_{7}\psi_{4}\bar{\psi}_{3}\gamma^{\mu\nu\lambda}\gamma_{7}\psi_{2}\right].\label{Fierz}\end{align}
where we consider commuting spinors.

Using $\bar{\psi}_{1}=\bar{\varepsilon_{+}}\gamma_{\mu}$, $\psi_{2}=\varepsilon_{+}$,
$\bar{\psi_{3}}=\bar{\varepsilon}_{+}$ and $\psi_{4}=\gamma_{\mu}\varepsilon_{+}$
in \eqref{Fierz} we obtain\begin{equation}
l_{\mu}^{+}l^{+\mu}=0\label{l+l+}\end{equation}
where we used the equations\begin{align*}
\gamma_{\mu}\gamma_{\nu}\gamma^{\mu} & =-4\gamma_{\nu}\\
\gamma_{\mu}\gamma_{\mu\nu\lambda}\gamma^{\mu} & =0.\end{align*}
In a similar manner one can obtain the relation\begin{equation}
l_{\mu}^{-}l^{-\mu}=0.\label{l-l-}\end{equation}
Using $\bar{\psi}_{1}=\bar{\varepsilon_{+}}$, $\psi_{2}=\varepsilon_{-}$,
$\bar{\psi_{3}}=\bar{\varepsilon}_{-}$ and $\psi_{4}=\varepsilon_{+}$
we obtain\begin{equation}
Z^{+}Z^{-}=\frac{1}{4}l_{\mu}^{+}l^{-\mu}+\frac{1}{48}q^{+\pi\rho\sigma}q_{\pi\rho\sigma}^{-}.\label{ZZ}\end{equation}
Choosing $\bar{\psi}_{1}=\bar{\varepsilon_{+}}\gamma^{\mu}$, $\psi_{2}=\varepsilon_{+}$,
$\bar{\psi_{3}}=\imath\bar{\varepsilon}_{+}$ and $\psi_{4}=\gamma_{\mu\nu\lambda}\varepsilon_{+}$we
have\begin{equation}
l^{+\mu}q_{\mu\nu\lambda}^{+}=-l^{+\mu}q_{\mu\nu\lambda}^{+}\Rightarrow i_{l^{+}}q^{+}=0.\label{l+q+}\end{equation}
In a similar way one obtains\begin{equation}
l^{-\mu}q_{\mu\nu\lambda}^{-}=-l^{-\mu}q_{\mu\nu\lambda}^{-}\Rightarrow i_{l^{-}}q^{-}=0.\label{l-q-}\end{equation}

Where we used the relations\begin{align*}
\gamma^{\mu}\gamma_{\rho}\gamma_{\mu\nu\lambda} & =4g_{\lambda\rho}\gamma_{\nu}-4g_{\nu\rho}\gamma_{\lambda}-2\gamma_{\nu\lambda\rho}\\
\gamma^{\mu}\gamma_{\rho\sigma\tau}\gamma_{\mu\nu\lambda} & =-12\delta_{\left[\sigma\tau\right.}^{\alpha\beta}\gamma_{\left.\rho\right]}g_{\alpha\lambda}g_{\beta\nu}-2\gamma_{\lambda\nu\rho\sigma\tau}.\end{align*}
We consider \eqref{Fierz} with $\bar{\psi}_{1}=\bar{\varepsilon}_{\pm}\gamma^{\mu\nu}\gamma^{\alpha}$,
$\psi_{2}=\varepsilon_{\pm}$, $\bar{\psi}_{3}=\bar{\varepsilon}_{\pm}$
and $\psi_{4}=\gamma_{\alpha}\gamma^{\gamma\delta}\varepsilon_{\pm}$
which gives\begin{align}
4\,\bar{\varepsilon}_{\pm}\gamma^{\mu\nu}\gamma^{\alpha}\varepsilon_{\pm}\bar{\varepsilon}_{\pm}\gamma_{\alpha}\gamma^{\gamma\delta}\varepsilon_{\pm} & =\bar{\varepsilon}_{\pm}\gamma^{\mu\nu}\gamma^{\alpha}\gamma_{\rho}\gamma_{\alpha}\gamma^{\gamma\delta}\varepsilon_{\pm}\bar{\varepsilon}_{\pm}\gamma^{\rho}\varepsilon_{\pm}-\frac{1}{12}\bar{\varepsilon}_{\pm}\gamma^{\mu\nu}\gamma^{\alpha}\gamma_{\rho\sigma\tau}\gamma_{\alpha}\gamma^{\gamma\delta}\varepsilon_{\pm}\bar{\varepsilon}_{\pm}\gamma^{\rho\sigma\tau}\varepsilon_{\pm}\nonumber \\
 & =-4\bar{\varepsilon}_{\pm}\gamma^{\mu\nu}\gamma_{\rho}\gamma^{\gamma\delta}\varepsilon_{\pm}\bar{\varepsilon}_{\pm}\gamma^{\rho}\varepsilon_{\pm}.\label{IIpr}\end{align}
We also list the identities\begin{align}
\gamma_{\mu\nu}\gamma_{\alpha} & =g_{\alpha\nu}\gamma_{\mu}-g_{\alpha\mu}\gamma_{\nu}+\gamma_{\alpha\mu\nu}\label{id1}\\
\gamma_{\alpha}\gamma_{\gamma\delta} & =g_{\alpha\gamma}\gamma_{\delta}-g_{\alpha\delta}\gamma_{\gamma}+\gamma_{\alpha\gamma\delta}\label{id2}\\
\gamma_{\mu\nu}\gamma_{\alpha}\gamma_{\gamma\delta} & =\gamma_{\alpha\gamma\delta\mu\nu}-6g_{\alpha\pi}g_{\gamma\rho}g_{\delta\sigma}\delta_{\:\mu\nu}^{\left[\pi\rho\right.}\gamma^{\left.\sigma\right]}+4g_{\alpha\pi}g_{\mu\rho}g_{\nu\sigma}\delta_{\:\gamma\delta}^{\pi\left[\rho\right.}\gamma^{\left.\sigma\right]}\label{id3}\\
 & \quad+6g_{\alpha\pi}g_{\gamma\rho}g_{\delta\sigma}\delta_{\left[\nu\right.}^{\left[\pi\right.}\gamma_{\quad\:\left.\mu\right]}^{\left.\rho\sigma\right]}+2g_{\alpha\left[\gamma\right.}\gamma_{\left.\delta\right]\mu\nu}\nonumber \end{align}
We now look at the Fierz identity involving $\bar{\psi}_{1}=\bar{\varepsilon}_{+}\gamma^{\gamma\delta}\gamma_{\mu}$,
$\psi_{2}=\varepsilon_{+}$, $\bar{\psi}_{3}=\bar{\varepsilon}_{-}$
and $\psi_{4}=\gamma_{\nu}\gamma_{\gamma\delta}\varepsilon_{-}$ which
gives after antisymmetrization in $\mu$ and $\nu$\begin{align}
4\;\bar{\varepsilon}_{+}\gamma^{\gamma\delta}\gamma_{\left[\mu\right.}\varepsilon_{+}\;\bar{\varepsilon}_{-}\gamma_{\left.\nu\right]}\gamma_{\gamma\delta}\varepsilon_{-} & =-12\bar{\varepsilon}_{+}\gamma_{\mu\nu}\varepsilon_{-}\;\bar{\varepsilon}_{-}\varepsilon_{+}+12\bar{\varepsilon}_{+}\varepsilon_{-}\;\bar{\varepsilon}_{-}\gamma_{\mu\nu}\varepsilon_{+}\nonumber \\
 & \quad-2\bar{\varepsilon}_{+}\gamma_{\mu\nu\rho\sigma}\varepsilon_{-}\;\bar{\varepsilon}_{-}\gamma^{\rho\sigma}\varepsilon_{+},\label{double}\end{align}
considering now the Fierz identity for $\bar{\psi}_{1}=\bar{\varepsilon}_{+}\gamma_{\nu}$,
$\psi_{2}=\varepsilon_{+}$, $\bar{\psi}_{3}=\bar{\varepsilon}_{-}$
and $\psi_{4}=\gamma_{\mu}\varepsilon_{-}$we have\[
\bar{\varepsilon}_{+}\gamma_{\mu\nu\rho\sigma}\varepsilon_{-}\;\bar{\varepsilon}_{-}\gamma^{\rho\sigma}\varepsilon_{+},=8\bar{\varepsilon}_{+}\gamma_{\left[\nu\right.}\varepsilon_{+}\bar{\varepsilon}_{-}\gamma_{\left.\mu\right]}\varepsilon_{-}+2\bar{\varepsilon}_{+}\gamma_{\mu\nu}\varepsilon_{-}\;\bar{\varepsilon}_{-}\varepsilon_{+}-2\bar{\varepsilon}_{+}\varepsilon_{-}\;\bar{\varepsilon}_{-}\gamma_{\mu\nu}\varepsilon_{+}.\]
Finally \eqref{double} takes the form\begin{equation}
\bar{\varepsilon}_{+}\gamma^{\gamma\delta}\gamma_{\left[\mu\right.}\varepsilon_{+}\;\bar{\varepsilon}_{-}\gamma_{\left.\nu\right]}\gamma_{\gamma\delta}\varepsilon_{-}=-4\bar{\varepsilon}_{+}\gamma_{\mu\nu}\varepsilon_{-}\;\bar{\varepsilon}_{-}\varepsilon_{+}+4\bar{\varepsilon}_{+}\varepsilon_{-}\;\bar{\varepsilon}_{-}\gamma_{\mu\nu}\varepsilon_{+}-4\bar{\varepsilon}_{+}\gamma_{\left[\nu\right.}\varepsilon_{+}\bar{\varepsilon}_{-}\gamma_{\left.\mu\right]}\varepsilon_{-}.\label{opa}\end{equation}
Another useful identity is generated by using the choice $\bar{\psi}_{1}=\bar{\varepsilon}_{+}\gamma_{\mu}$,
$\psi_{2}=\varepsilon_{+}$, $\bar{\psi}_{3}=\bar{\varepsilon}_{+}$
and $\psi_{4}=\gamma^{\mu}\gamma_{\nu}\varepsilon_{-}$ which leads
to \begin{align}
\bar{\varepsilon}_{+}\gamma_{\mu}\varepsilon_{+}\bar{\varepsilon}_{+}\gamma^{\mu}\gamma_{\nu}\varepsilon_{-} & =0\Rightarrow\nonumber \\
\bar{\varepsilon}_{+}\gamma^{\mu}\varepsilon_{+}\bar{\varepsilon}_{+}\gamma_{\mu\nu}\varepsilon_{-} & =-\bar{\varepsilon}_{+}\varepsilon_{-}\bar{\varepsilon}_{+}\gamma_{\mu}\varepsilon_{+}\nonumber \\
l^{+\mu} & U_{\mu\nu}^{+}=-Z^{+}l_{\nu}^{+}.\label{l+U+}\end{align}
In a similar way one may also prove that\begin{equation}
l^{-\mu}U_{\mu\nu}^{-}=-Z^{-}l_{\nu}^{-}.\label{l-U-}\end{equation}
Using \eqref{Idef},\eqref{Jdef} and \eqref{IJ} in \eqref{opa}
and after contracting with $\bar{\varepsilon}_{+}\gamma_{\mu}\varepsilon_{+}$
we can prove with the help of \eqref{l+U+} that\begin{equation}
l^{+\mu}U_{\mu\nu}^{-}=Z^{-}l_{\nu}^{+}\label{l+U-}\end{equation}
and in a similar way \begin{equation}
l^{-\mu}U_{\mu\nu}^{+}=Z^{+}l_{\nu}^{-}.\label{l-U+}\end{equation}

\section{Algebraic equations for the bilinears}

\subsubsection*{Scalar Identities}

Multiplying \eqref{kilS3} by $\imath\bar{\varepsilon}\gamma_{7}$
and \eqref{kilS1} by $\bar{\varepsilon}$ we obtain the relations\begin{align*}
-\frac{\alpha}{2}e^{-\frac{1}{2}\left(H+G\right)}f_{1}-\frac{1}{4}L^{\mu}\partial_{\mu}\left(H+G\right)-\frac{1}{4}e^{-\frac{3}{2}\left(G+H\right)}Y^{\mu\nu}F_{\mu\nu} & =0\\
\frac{n}{2}e^{-\frac{1}{2}\left(H-G\right)}f_{2}-\frac{1}{4}L^{\mu}\partial_{\mu}\left(H-G\right)+\frac{1}{4}e^{-\frac{3}{2}\left(G+H\right)}Y^{\mu\nu}F_{\mu\nu} & =0.\end{align*}
If we now multiply \eqref{kilS3} by $\bar{\varepsilon}$and \eqref{kilS1}
by $\imath\bar{\varepsilon}\gamma_{7}$ and equate the real and imaginary
parts of the corresponding equations to zero we the relations\begin{align}
K^{\mu}\partial_{\mu}\left(H+G\right) & =0\label{Kp(H+G)}\\
2\alpha f_{2} & =e^{-\left(G+H\right)}V^{\mu\nu}F_{\mu\nu}\\
K^{\mu}\partial_{\mu}\left(H-G\right) & =0\label{Kp(H-G)}\\
-2f_{1}n & =e^{-\left(H+2G\right)}V^{\mu\nu}F_{\mu\nu}.\end{align}

\subsubsection*{Vector Identities}

We now consider \eqref{kilS3} and its conjugate again but this time
we multiply by $-\imath\bar{\varepsilon}\gamma_{\mu}\gamma_{7}$ and
$-i\gamma_{\mu}\gamma_{7}\varepsilon$ respectively and add them.
The result of the operation reads\begin{equation}
\alpha L_{\mu}e^{-\frac{1}{2}\left(G+H\right)}+\frac{1}{2}f_{1}\partial_{\mu}\left(G+H\right)+\frac{1}{2}\star\left(\Omega\wedge F\right)_{\mu}e^{-\frac{3}{2}\left(G+H\right)}=0.\label{v1}\end{equation}
Multiplying \eqref{kilS1} and its conjugate by $\imath\bar{\varepsilon}\gamma_{\mu}\gamma_{7}$
respectively $\imath\gamma_{\mu}\gamma_{7}\varepsilon$ and adding
the resulting equations we obtain\begin{equation}
-ne^{-\frac{1}{2}\left(H-G\right)}L_{\mu}+\frac{1}{2}f_{2}\partial_{\mu}\left(H-G\right)-\frac{1}{2}e^{-\frac{3}{2}\left(G+H\right)}F_{\mu}^{\:\lambda}K_{\lambda}=0.\label{v2}\end{equation}
We now turn to \eqref{kilS1} and we multiply it by $\bar{\varepsilon}\gamma_{\mu}$giving
back\begin{align}
-f_{1}\partial_{\mu}\left(H-G\right)+\star\left(F\wedge\Omega\right)_{\mu}e^{-\frac{3}{2}\left(G+H\right)} & =0.\label{v3}\end{align}
Multiplying \eqref{kilS3} by $\bar{\varepsilon}\gamma_{\mu}$we obtain\begin{align}
f_{2}\partial_{\mu}\left(H+G\right)-e^{-\frac{3}{2}\left(G+H\right)}F_{\mu}^{\:\lambda}K_{\lambda} & =0\label{v4}\end{align}

\subsubsection*{Rank two identities}

Multiplying \eqref{kilS3} and \eqref{kilS1} by $\bar{\varepsilon}\gamma_{\mu\nu}$
and $\bar{\varepsilon}\gamma_{\mu\nu}\gamma_{7}$, considering the
identities\begin{align*}
\gamma_{\mu\nu}\gamma_{\lambda} & =2g_{\lambda\left[\nu\right.}\gamma_{\left.\mu\right]}+\gamma_{\mu\nu\lambda}\\
\gamma_{\lambda}\gamma_{\mu\nu} & =-2g_{\lambda\left[\nu\right.}\gamma_{\left.\mu\right]}+\gamma_{\mu\nu\lambda}\\
\gamma_{\mu\nu}\gamma_{\kappa\lambda} & =-2g_{\mu\left[\kappa\right.}g_{\left.\lambda\right]\nu}+\gamma_{\mu\nu\kappa\lambda}+2\gamma_{\kappa\left[\mu\right.}g_{\left.\nu\right]\lambda}-2\gamma_{\lambda\left[\mu\right.}g_{\left.\nu\right]\kappa}\end{align*}
and taking separately the real and imaginary parts to zero we have\begin{align}
\frac{\alpha}{2}e^{-\frac{1}{2}\left(H+G\right)}Y_{\mu\nu}-\frac{1}{24}\epsilon_{\mu\nu\pi\rho\sigma\lambda}\Omega^{\pi\rho\sigma}\partial^{\lambda}\left(H+G\right)+\frac{1}{2}f_{1}e^{-\frac{3}{2}\left(G+H\right)}F_{\mu\nu}+\frac{1}{8}\epsilon_{\mu\nu\kappa\lambda\pi\rho}F^{\kappa\lambda}V^{\pi\rho} & =0\\
L_{\left[\mu\right.}\partial_{\left.\nu\right]}\left(H+G\right)+e^{-\frac{3}{2}\left(G+H\right)}Y_{\kappa\left[\mu\right.}F_{\left.\nu\right]}^{\quad\kappa} & =0\end{align}

\begin{align}
\frac{\alpha}{2}V_{\mu\nu}e^{-\frac{1}{2}\left(H+G\right)}-\frac{1}{4}K_{\left[\mu\right.}\partial_{\left.\nu\right]}\left(G+H\right)-\frac{1}{2}e^{-\frac{3}{2}\left(G+H\right)}F_{\mu\nu}f_{2}-\frac{1}{8}\epsilon_{\kappa\lambda\mu\nu\pi\rho}Y^{\pi\rho}F^{\kappa\lambda} & e^{-\frac{3}{2}\left(G+H\right)}=0\\
\frac{1}{4}\Omega_{\mu\nu}^{\quad\:\lambda}\partial_{\lambda}\left(G+H\right)-V_{\kappa\left[\mu\right.}F_{\left.\nu\right]}^{\quad\kappa}e^{-\frac{3}{2}\left(G+H\right)} & =0\label{omega1}\end{align}

\begin{align}
\frac{n}{2}e^{-\frac{1}{2}\left(H-G\right)}V_{\mu\nu}+\frac{1}{24}\epsilon_{\mu\nu\lambda\pi\rho\sigma}\partial^{\lambda}\left(H-G\right)\Omega^{\pi\rho\sigma}-\frac{1}{2}F_{\mu\nu}e^{-\frac{3}{2}\left(G+H\right)}f_{1}-\frac{1}{8}\epsilon_{\mu\nu\kappa\lambda\pi\rho}F^{\kappa\lambda}V^{\pi\rho}e^{-\frac{3}{2}\left(G+H\right)} & =0\\
L_{\left[\nu\right.}\partial_{\left.\mu\right]}\left(H-G\right)+e^{-\frac{3}{2}\left(G+H\right)}Y_{\kappa\left[\mu\right.}F_{\left.\nu\right]}^{\quad\kappa} & =0\end{align}

\begin{align}
\frac{n}{2}e^{-\frac{1}{2}\left(H-G\right)}Y_{\mu\nu}+\frac{1}{4}K_{\left[\nu\right.}\partial_{\left.\mu\right]}\left(H-G\right)-\frac{1}{2}F_{\mu\nu}f_{2}e^{-\frac{3}{2}\left(G+H\right)}-\frac{1}{8}\epsilon_{\mu\nu\kappa\lambda\pi\rho}Y^{\pi\rho}F^{\kappa\lambda}e^{-\frac{3}{2}\left(G+H\right)} & =0\\
\frac{1}{4}\Omega_{\mu\nu\lambda}\partial^{\lambda}\left(H-G\right)+V_{\kappa\left[\mu\right.}F_{\left.\nu\right]}^{\quad\kappa}e^{-\frac{3}{2}\left(G+H\right)} & =0.\label{omega2}\end{align}

\subsubsection*{Rank Three Identities}

We now multiply \eqref{kil1} by $\bar{\varepsilon}\gamma_{\mu\nu\kappa}$and
take the real and imaginary part separately\begin{align}
\alpha e^{-\frac{1}{2}\left(G+H\right)}\Omega_{\mu\nu\kappa}+\frac{1}{2}\epsilon_{\mu\nu\kappa\rho\alpha\beta}Y^{\alpha\beta}\partial^{\rho}H & =0\nonumber \\
ne^{-\frac{1}{2}\left(H-G\right)}\star\Omega_{\mu\nu\kappa}+3\partial_{\left[\kappa\right.}HV_{\left.\mu\nu\right]} & =0.\label{Omega_wedge_L}\end{align}
Doing the same with \eqref{kil2} we obtain the equations\begin{align}
\alpha e^{-\frac{1}{2}\left(G+H\right)}\Omega_{\mu\nu\kappa}+\frac{1}{2}\epsilon_{\mu\nu\kappa\rho\alpha\beta}Y^{\alpha\beta}\partial^{\rho}G+6e^{-\frac{3}{2}\left(G+H\right)}K_{\left[\kappa\right.}F_{\left.\mu\nu\right]}+e^{-\frac{3}{2}\left(G+H\right)}\epsilon_{\kappa\mu\nu\pi\rho\sigma}F^{\pi\rho}L^{\sigma} & =0\label{rank3eq}\\
ne^{-\frac{1}{2}\left(H-G\right)}\star\Omega_{\mu\nu\kappa}-3\partial_{\left[\kappa\right.}GV_{\left.\mu\nu\right]}-2e^{-\frac{3}{2}\left(G+H\right)}F_{\quad\left[\nu\right.}^{\rho}\Omega_{\left.\kappa\mu\right]\rho} & =0.\nonumber \end{align}
\newpage{}

\bibliographystyle{utphys}
\bibliography{Test}

\end{document}